\documentclass[12pt, letterpaper]{article}

\usepackage{amsmath}
\usepackage{amsfonts}
\usepackage{amssymb}
\usepackage{graphicx}

\usepackage{color}

\usepackage{amsmath, amssymb, graphics, epsfig, graphicx}
\usepackage{epsf}
\usepackage{epstopdf}
\usepackage {amssymb}
\newcommand{\nc}{\newcommand}
\nc{\ba}{\begin{eqnarray}}
\nc{\ea}{\end{eqnarray}}
\newcommand\be{\begin{equation}}
\newcommand\ee{\end{equation}}

\newcommand{\calR}{{\cal{R}}}

\newcommand{\calP}{{\cal{P}}}

\newcommand{\bea}{\begin{eqnarray}}
\newcommand{\eea}{\end{eqnarray}}

\begin{document}

\vspace{5mm}
\vspace{0.5cm}
\begin{center}

\def\thefootnote{\fnsymbol{footnote}}

{\Large  Primordial Universe Inside the Black Hole and  Inflation }
\\[0.5cm]

{  Hassan Firouzjahi}
\\[0.2cm]
{\small \textit{School of Astronomy, Institute for Research in Fundamental Sciences (IPM) \\ P.~O.~Box 19395-5531, Tehran, Iran
}}\\

\vspace{0.1cm}
{\small {
 firouz@ipm.ir
}}\\

\end{center}

\vspace{.8cm}

\hrule \vspace{0.3cm}


\begin{abstract}

We speculate that the early Universe was inside a primordial black hole. The interior of the the black hole is 
a dS background and the two spacetimes are separated on the surface of black hole's event horizon. 
We argue that this picture provides a natural realization of  inflation without invoking the inflaton field. The black hole evaporation by  Hawking radiation provides a natural mechanism for terminating inflation so reheating and the hot big bang cosmology starts from the evaporation of  black hole to relativistic particles. The quantum gravitational fluctuations at the boundary of black hole generate the nearly 
scale invariant  scalar and tensor perturbations with the ratio of tensor to scalar power spectra at the order  of $10^{-3}$.  
As the black hole evaporates, the radius of its event horizon shrinks and the Hubble expansion rate during inflation increases slowly so the quantum Hawking radiation provides a novel mechanism for
the violation of null energy condition in cosmology.

\end{abstract}
\vspace{0.5cm} \hrule
\def\thefootnote{\arabic{footnote}}
\setcounter{footnote}{0}
\newpage
\section{Introduction}

Inflation is a cornerstone of the theories of primordial Universe \cite{Guth:1980zm, Linde:1981mu, Albrecht:1982wi, Sato:1980yn, Starobinsky:1980te}.  The basic predictions of inflation that the perturbations on CMB maps to be nearly scale invariant, nearly adiabatic and nearly Gaussian are very well consistent with observations \cite{Ade:2015lrj, Ade:2015xua, Planck:2013jfk}. 

Despite its immense successes with observations, there is no unique realization of inflation dynamics. 
There are numerous models of inflation which are well consistent with data. In its simplest form, the mechanism of inflation is based on a scalar field, the inflaton field, which slowly rolls on the top of its nearly
flat potential.   The quantum fluctuations of the light inflaton field generate the  perturbations 
which are stretched on super-Hubble scales to seed the perturbations on CMB and large scale structure.
Despite its appealing simplicity, this picture suffers from some important shortcomings. One important question is 
the nature of inflaton field, i.e. what field in the theories of high energy physics, say beyond Standard Model (SM),  plays the role of inflaton field. The second important question is the vast diversity of models which are nearly degenerate with the current observations. Another important question is the initial singularity associated with big bang cosmology. In a sense inflation only pushes the big bang singularity to initial times in evolution of Universe hoping that it is beyond the reach of observational cosmology. 
Besides these conceptual questions, there are technical questions such as what mechanism, i.e. symmetry principle, keeps the inflation field light during a long enough period  of inflation. Another question is how the SM degrees of freedoms are generated during (p)reheating. The current models of (p)reheating are far from providing a detailed mechanism to generate SM fields. Motivated by these questions, it is important to think if inflation is unique in our primordial Universe. One natural question is if there are alternatives to inflation which possess all the successes of inflation concerning observations and yet bypass the above mentioned shortcomings.   This line of thought has been pursued in some models of alternative to inflation, for example see \cite{Khoury:2001wf, Steinhardt:2001st, Ijjas:2013vea, Brandenberger:1988aj, Nayeri:2005ck}. 

In this work, we speculate whether the interior of a black hole can be a place to look for primordial Universe either to realize inflation or  its alternative.  Indeed, the recent detections of gravitational wave by the LIGO team \cite{Abbott:2016blz, Abbott:2016nmj} from the merging of two binary black holes 
have put the reality of black hole beyond doubt. Massive and supermassive black holes are ubiquitous in the cosmos which are detected indirectly by astronomical observations. Similarly, primordial black holes are expected to exist in early Universe. Black holes emit black body radiation via 
Hawking mechanism \cite{Hawking:1974sw} and the lifetime of a black hole roughly scales with its mass as $M^3$. Therefore, primordial black hole of mass smaller than $10^{14}$ grams are believed to evaporate via Hawking radiation. 
 
Black hole physics play a key role in understanding quantum gravity which was the subject of  extensive studies in the past decades. The connection of  black hole physics to thermodynamics and the fact that black hole emits thermal radiation are very curious \cite{Hawking:1974sw, Hartle:1976tp}. In addition, there have been attempts to provide a link between the temperature of black hole associated with its horizon to the temperature of dS horizon \cite{Gibbons:1977mu}. This is motivated from the fact that  both the black hole and the dS backgrounds have horizons (though different in nature) and the temperature of the black hole is associated with its horizon. The fact that the associated temperature of the dS background,   $T_{dS}= H/2\pi$,  is the same as the amplitude of quantum fluctuations of massless scalar fields in dS background makes this connection intriguing.

In this work, we speculate whether the interior of a black hole could be the host of our primordial Universe and a new mechanism of inflation can be realized. 


\section{Cosmology Inside a Black Hole}
\label{setup}

In this section we present our setup. We speculate that the primordial Universe is inside a black hole. 
The interior cosmological background are separated from the outside black hole solution by the black hole's event horizon.  Not much is known for the physics inside  the event horizon. The simple reason is that  the interior  region is causally disconnected from the outside region, for a review of black hole physics see \cite{Padmanabhan}.  Any in-falling signal smoothly passes through the event horizon while no signal can escape past the event horizon. Motivated by this curious phenomena, we postulate that the interior of black hole is a cosmological background bounded by cosmological horizon. 
This intuition is supported from the fact that inside the horizon, the role of $ t$ and $  r$ as the time-like and space-like coordinates have switched. In this picture we speculate that 
the Hubble horizon  (cosmological horizon) from inside is attached to the event horizon  from the outside.

Indeed, the idea of replacing the interior of the black hole by a dS spacetime is not new. There have been several works in the past suggesting that the interior of black hole may be replaced by a non-singular dS geometry  \cite{Sakharov:1966aja, Poisson:1988wc, Frolov:1989pf, Frolov:1988vj, Shen-Zhu}. For similar ideas but is somewhat different contexts see also \cite{Sato:1981bf, Maeda:1981gw, Sato:1981gv, Farhi:1986ty, Blau:1986cw, Oshita:2016btk}. 
One motivation in replacing the interior of black hole by the dS background was to get away with the singularity of black hole. It is generally believed that the singularity of black hole is a shortcoming of the classical  general relativity. On very small scales, say on Planck scale $\ell_P \equiv  \sqrt{G}$ with $G$ being Newton's constant,   it is expected that the quantum effects become important which provide mechanisms to resolve the singularity inside the black hole. This expectation brings one to the realms of quantum gravity which is not understood at the moment.  

In particular in  \cite{Poisson:1988wc} it is argued that the gravitational vacuum polarization inside the black hole acts as a mechanism of self-regularization. In this picture the Schwarzschild geometry with the mass $M$ is valid down to the quantum barrier radius (in Planck unit)  $r= r_Q = (M)^{1/3}$  where the curvature of spacetime $R \sim M/r^3 $  grows to order unity. 
Below this radius, down to radius where the  quantum gravity effects become strong, $r_{QG} \lesssim r < r_Q$, the geometry is nearly classical described by the classical Einstein equation  
$G_{\mu \nu} = 8 \pi T_{\mu \nu} ( \mathrm{vacuum \, polarization) }  \sim \mathrm{constant}$, describing a dS spacetime.

The proposal that there is an upper bound on the curvature of spacetime \cite{Markov} has been employed in 
\cite{Frolov:1989pf, Frolov:1988vj} to replace the interior of black hole by the dS spacetime. It is argued that the divergence of the spacetime curvature at the center of black hole is an artifact of classical Einstein equations. Taking the quantum effects into account, there is a universal upper bound to the curvature of spacetime preventing the appearance of singularity. Naturally one expects that this universal length scale, $\ell_c$,  to be near the Planck length, $\ell_c \sim \ell_P$. Of course, similar to proposal  in \cite{Poisson:1988wc}, this proposal is put forwarded as a hypothesis waiting for a final theory of quantum gravity for justification.  Below we summarize the relevant results   in \cite{Frolov:1989pf, Frolov:1988vj} which are useful  for our setup. 

The existence of limiting curvature for a spherically symmetric spacetime is postulated in terms of the Riemann tensor as 
\ba\label{max-curvature}
R_{\alpha \beta \gamma \delta } R^{\alpha \beta \gamma \delta} \lesssim \alpha_c \ell_c^{-4}
\ea
 in which $\alpha_c$ is a dimensionless parameter. This condition also implies that other quadratic curvature invariants constructed from Ricci tensor and Weyl tensor are finite.  As a second assumption, it is assumed that when the curvature reaches its maximum value the equation of state approaches the vacuum (dS) type  
\ba
\label{dS-vac}
R^{\mu}_{\,  \nu} =3\ell_c^{-2} \delta^\mu_\nu \, .
\ea 
As a motivation for this assumption, note that the spacetime inside the black hole is homogeneous but anisotropic. This anisotropy is understood from the fact that inside the black hole's horizon the role of $t$ and $r$ as the time-like and space-like coordinates are switched. Therefore,  any dependence on the variable $r$ inside the horizon is interpreted as a function of time for the observer inside the black hole. Consequently, the interior of black hole is similar to  Kasner spacetime describing anisotropic contractions.  This anisotropic behavior is a consequence of classical Einstein's field equations. However, once the quantum effects such as the vacuum polarizations  \cite{Poisson:1988wc} are taken into account, the anisotropy is damped while the curvature tensor reaches its  maximum  value near the Planck scale. Intuitively speaking, the vacuum equation of state Eq. (\ref{dS-vac}) is the simplest case which arise from this prescription and yet regularizing the curvature singularity. 

Imposing the spherical symmetry, the black hole metric is given by the  Schwarzschild solution 
\ba
\label{BH-metric1}
ds^2 = - f( r) d  {\bar t} ^{\,2} + f( r)^{-1} d  r^2 +  r^2 d \Omega^2    \, ,
\ea
in which $ r$ is the radial coordinate, $d \Omega^2$ represents the geometry of  the two sphere and 
\ba
\label{f-Schw}
f( r) = 1- \frac{2 M}{ r} \, ,
\ea
in which $M$ is the mass of black hole (in Planck unit)  as measured by an observer very far from black hole.  The black hole event horizon is located at $r_S \equiv 2  M$. Note that in order not to confuse with the cosmic time $t$, we denote the time coordinate in black hole solution by $\bar t$. 

Imposing the maximum universal curvature hypothesis Eq. (\ref{max-curvature}) and the vacuum assumption (\ref{dS-vac}), the  interior of black hole is glued to a dS space at the position $r_0 < r_S$,  where 
\ba
r_0 =  2^{1/6} H^{-1} ( M/H)^{1/3} \, .
\ea
For the region $r< r_0$, the spacetime is in dS type with the metric in  static coordinate given by 
Eq. (\ref{BH-metric1}) with $f(r) = 1- (r H)^2$ in which  $H^{-1}$ represents the radius of dS horizon. 

\begin{figure}
  \centering
  \includegraphics[width=0.7\textwidth]{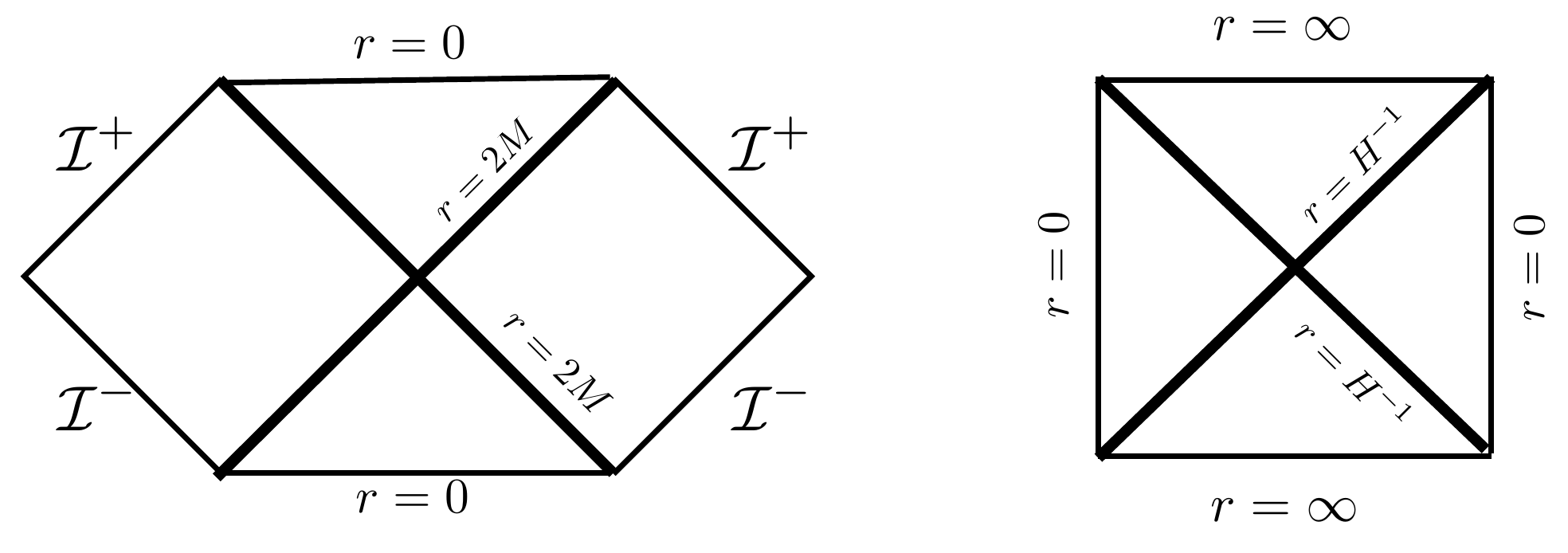} 
  \caption{The conformal diagrams: (left) black hole, (right) dS space.  In both figures the horizons are denoted by the thick lines. 
  }
  \label{bh-ds-fig1}
\end{figure}

For the above picture to be a consistent solution of the Einstein's field equations, one has to impose the 
appropriate matching conditions \cite{Israel:1966rt}. More specifically, denoting the surface of matching 
at $r=r_0$ by the three-dimensional $\Sigma$, then the intrinsic metric induced on $\Sigma$ from both dS and black hole sides should be continuous while the jump in extrinsic curvature $[ K^m_n] $ is balanced by the surface energy density $S^m_n$ via Israel junction condition $[ K^m_n] - \delta^m_n [ K^m_m] = 8 \pi S^m_n$. Imposing  these matching condition one can find the tension and pressure on the surface $\Sigma$ in order to support the above solution \cite{Frolov:1988vj}.

\begin{figure}[t]
  \vspace{-0.5cm}
  \centering
  \includegraphics[width=0.4\textwidth]{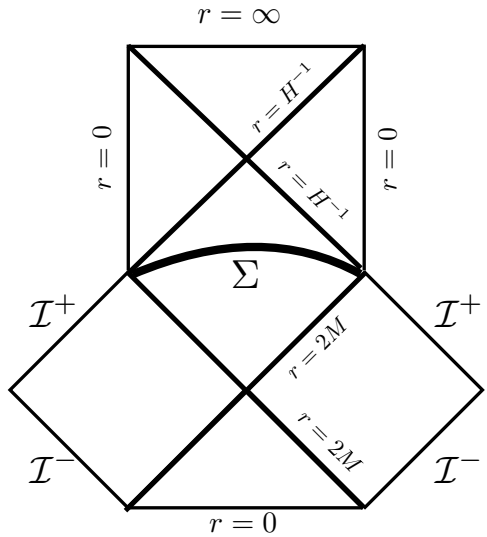} 
  \vspace{-1cm}
  \caption{The conformal diagram when the dS space is glued to the black hole interior on the space-like surface $\Sigma$ at $r=r_0$.   }
  \label{bh-ds-fig2}
\end{figure}

It is important to note that the surface of matching $\Sigma$ at $r=r_0$ is a space-like surface since the coordinate  $r$ inside the black hole's event horizon is a time-like coordinate.  The surface $\Sigma$ has the topology of $S^2 \times R^1$ which is a tube with the fixed radius equal to $r_0$ extended 
along the direction of  $\bar t$.

\begin{figure}[t]
  \vspace{-2cm}
  \centering
  \includegraphics[width=0.6\textwidth]{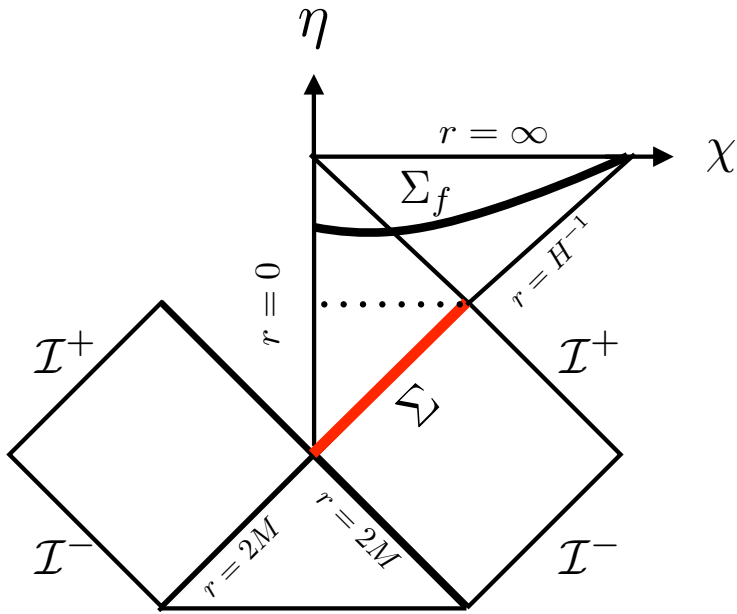} 
  \vspace{-2cm}
  \caption{The conformal diagram of our setup when the dS horizon is glued to the black hole's event horizon.  The surface of matching $\Sigma$,  located at $r= 2 M = H^{-1}$,  is shown by the thick red line. The dotted line represents the surface when deflation ends at $\eta =-\pi/2$. The curve 
  $\Sigma_f$ represents the surface of end of inflation.  After inflation ends on the surface $\Sigma_f$, the Universe enters the standard hot big bang phase.  The  coordinates ranges in dS space are $-\pi < \eta <0 $ and $\pi< \chi <0$.    }
  \label{setup-fig}
\end{figure}

The conformal Penrose diagrams associated for the above discussions are presented in Fig. \ref{bh-ds-fig1} and Fig. \ref{bh-ds-fig2}.  In Fig. \ref{bh-ds-fig1} the conformal diagram for the full black hole and dS spaces before joining  are plotted.  In Fig. \ref{bh-ds-fig2}  the conformal diagram where the interior of black hole for the region $r<r_0$ is replaced by the dS space is presented. The matching surface 
$\Sigma $ is denoted with the thick curve in the middle.  As can be seen from this figure, the interior of black hole hosts a closed universe with positive spatial curvature.  The dS space for the times  after matching is a contracting deflation phase followed by an expanding inflationary stage  in which  the inflationary expansion continues forever.

The stability of the above model  was investigated in \cite{Balbinot:1990zz}. It was shown that the model is stable under small fluctuations meaning that the surface $\Sigma$ retains its $S^2\times R^1$ topology. The extension of the model of \cite{Frolov:1989pf, Frolov:1988vj} to the case where the interior of black hole is filled with a distribution of density profile (instead of sharp localization of density on a space-like surface) was presented in 
\cite{Dymnikova:1992ux} in which one replaces the cosmological constant $\Lambda$ with an $r$-dependent function \cite{Dymnikova:1999cz}.  

The setup we have in mind is similar to the model presented above. However, in order to make direct contact with the physics of black hole we make the further assumption that the matching surface $\Sigma$ is on the surface of black hole's event horizon with $r_0= r_S=  2 M = H^{-1}$. The conformal Penrose diagram for our setup is presented in Fig. \ref{setup-fig}. This has the appealing feature that the two spacetimes are glued along their corresponding horizons. As such, the matching surface $\Sigma$ is a null hypersurface. The conventional matching conditions \cite{Israel:1966rt} is not directly applicable when a null hypersurface separates the two spacetimes.   A general prescription when the two spacetimes are separated by a null surface is presented in \cite{Barrabes:1991ng}. It is shown that things are non-trivial in our setup. Indeed, to satisfy the matching conditions one encounters either divergence in surface pressure  or  kink on the null surface. This pathology is associated with the fact that  the continuity of pressure across the hypersurface is broken \cite{Poisson:1988wc}, see also \cite{Gron}.  In the Appendix we present the simple analysis of matching conditions to indicate the difficulty in joining the dS space to the  black hole along their horizons. 

The above discussions suggest that replacing the whole interior of black hole by a dS space is far from obvious. As just mentioned this requires an infinite surface pressure or the appearance of kinks. 
At the moment, we have no physical methods at hand to resolve these pathologies.  Having said this, we proceed with our phenomenological approach and assume that either by smearing the infinity in pressure on the surface of event horizon or by some quantum effects one can in principle glue the two spaces on the surface of horizons.  Assuming  the existence of a working matching condition   we proceed to investigate the cosmological implications of this setup.

As the conformal diagram Fig. \ref{setup-fig} suggests,  the dS space is a closed Universe with the metric
\ba
\label{ds1}
ds^2 = - dt^2 + H^{-2} \cosh^2(H\,  t) \left[  d \chi^2 + \sin^2 \chi d \Omega^2 
\right] ,
\ea
in which $t$ is the cosmic time with $ -\infty < t < + \infty $  and $0< \chi < \pi$.  Note that the cosmic time $t$ is different than the time coordinate in black hole solution which is denoted by $\bar t$.

Upon changing to conformal time $\eta$ via
\ba
\cosh(H\,  t) = -\frac{1}{\sin \eta} \, ,
\ea
the above metric is cast into
\ba
\label{ds2}
ds^2 = \frac{1}{H^2 \sin^2 \eta} \left(   -d \eta^2 + d \chi^2 + \sin^2 \chi d \Omega^2 \right) \, ,
\ea
in which $\eta$ changes in the interval $-\pi < \eta <0$. For a review of dS space in various coordinate systems see \cite{Mukhanov:2005sc}. 


We are interested in cosmological evolution  of the dS space inside the black hole. For the region 
$ -\pi < \eta <-\pi/2 $ in Fig. \ref{setup-fig} the Universe inside the black hole is in contracting deflationary phase.
This is clearly seen by the metric Eq. (\ref{ds2}). The Universe reaches its minimum size $H^{-1}$ at the point of bounce $\eta =-\pi/2$. This point is indicated by the dotted line in Fig. \ref{setup-fig}. Afterwards, the Universe enters the expanding inflationary phase continuing forever for $\eta \rightarrow 0$. 

Crucial to our discussions is the surface of end of inflation. In the current setup in which the interior of black hole is an exact  dS space inflation never end. Actually  the deflation phase  is eternal in the past 
$-\infty < t <0$ while the expanding inflationary  phase continues forever for $t>0$. Having said these, in Fig. \ref{setup-fig} we have plotted the curve $\Sigma_f$ as the surface of end of inflation. In next Section we discuss the physical mechanism which terminates inflation justifying the existence of final surface $\Sigma_f$.  It is understood that after inflation ends on the final surface $\Sigma_f$, the Universe enters the standard hot big bang phase which in this picture is  a closed radiation dominated Universe.

Before closing this Section let us look at the induced metric on the surface of matching $\Sigma$ where $r= r_S = 2 M = H^{-1}$.  In the coordinate system Eq. (\ref{ds1}) the surface of matching is given by
\ba
\label{matching-r}
a(t) \sin \chi = r_S = 2 M
\ea
where $a(t)$ is the scale factor in which for the metric Eq. (\ref{ds1}) it is $a(t) = H^{-1}\cosh (H\, t)$.

From the black hole side the induced metric on the null surface is simply $ds_{in}^2= r_S^2 d \Omega^2  $. From the dS side, the induced metric on the hypersurface of matching is
\ba
ds_{in}^2 = \left( -1 + \dot a^2\tan^2 \chi \right) dt^2 + a(t)^2 \sin \chi^2 d \Omega^2 \, ,
\ea 
in which a dot represents the derivative with respect to cosmic time $t$. The continuity of the induced metric on the matching surface simply requires that the coefficients of $dt^2$ above to vanish,  yielding
\ba
\left( \frac{\dot a}{a} \right)^2 + \frac{1}{a^2} = \frac{1}{r_S^2}  \, .
\ea
This is the Friedmann equation for a closed Universe with the scale factor $a(t) = H^{-1}\cosh (H\, t)$ and the  Hubble radius $H= r_S^{-1}$ as expected. This also suggests that the energy density of the Universe $\rho$ is 
\ba
\rho = \frac{3}{8 \pi G r_S^2} = \frac{3}{ 4\pi r_S^3} M \, .
\ea
This is well consistent with the intuition that the expansion of Universe is generated by the mass of black hole encompassed within the volume $4 \pi r_S^3/3$. 

So far we have considered the Schwarzschild solution. One can  extend these discussions  to the case of charged black hole with the 
Reissner-Nordstrom metric where
\ba
\label{RN-metric}
f(r)= 1- \frac{2  M}{ r} + \frac{Q^2}{ r^2} \, ,
\ea
in which $Q$ is the electric charge of the black hole. The condition that the singularity to be protected by the event horizon requires that $Q^2 \leq  M$, in which the radius of event horizon is given by
\ba
 r_+ = G M + (  M^2 - Q^2)^{1/2} \, .
\ea
Following the same steps as for the case of Schwarzschild metric and assuming that the black hole's event horizon coincides with the cosmological horizon, $r_+ = H^{-1}$,   we obtain 
\ba
\label{ds-sol1b}
\left( \frac{\dot a}{a} \right)^2 + \frac{1}{a^2} =  
\frac{ 2 G M }{ r_S^3}  - \frac{Q^2}{  r_S^4} \, .
\ea
As before, this describes a closed dS background  with the total energy density 
\ba
\label{rho-Q}
\rho = \frac{3}{4 \pi r_S^3} \left( M  - \frac{Q^2}{ 2 r_S} \right)  \, .
\ea
Interestingly, we see that the effective dS energy density receives a negative contributions from the charge $Q$.


\section{A New Mechanism of Inflation }
\label{inflation}

In the previous section we have speculated that the spacetime inside the event horizon of a black hole may be in the form of the dS geometry. As such, this provides a a natural  mechanism of inflation for early Universe. The novelty of this picture is that there is no need for inflaton field. Of course, before claiming that this
is a consistent realization of cosmic inflation, we have to answer two important questions. The first question is how inflation ends in this setup. In order to match to the observed hot big bang cosmology, inflation  has to end. The second question is   how perturbations are generated in this setup to seed the perturbations on CMB and large scale structure. Indeed, it is the predictions of any model of inflation for cosmological perturbations which bring the model in contact to observations.  

Below we answer each question in turn. To simplify the analysis we neglect the effect of spatial curvature as its effect is rapidly diluted during inflation.  Also we restore the Newton constant $G$ in the following analysis. 

\subsection{Background  }
\label{background}

Starting with $H= 1/ r_S$ and noting that $ r_S = 2 GM$ for Schwarzschild black hole, we obtain
\ba
\label{H-eq}
H = \frac{4 \pi M_P^2}{M} \, ,
\ea
in which $M_P^2\equiv 1/8 \pi G$ is the reduced Planck mass.  The above equation determines the
scale of inflation. In order for our classical treatment of the interior geometry of black hole to be valid, we require that $H \ll M_P$ so we can safely neglect the quantum gravity effects in dS background. 
This in turn implies that $M \gg M_P$ so the mass of our primordial black hole is much higher than the
scale of quantum gravity. This is also translated into $ r_S \gg \ell_P $ in which $\ell_P \equiv  \sqrt G \sim 1/M_P$. This means that on the black hole side, the classical general relativity  is trusted with quantum effects as small perturbations. 

In conventional models of inflation, inflation ends because of the slow-roll dynamics of the inflaton field. 
During inflation the slow-roll parameter $\epsilon= -\dot H/H^{2}$ is very small so the background geometry is very close to   a dS spacetime. As the inflaton field rolls towards its minimum, the deviation from the slow-roll condition
become significant and inflation ends eventually when $\epsilon =1$. We not that in all these models of inflation $\dot H <0$ as required from the weak energy condition, i.e. $\rho + P \ge 0$, so $\epsilon >0$
by construction. 

Looking at our expression for $H$ in Eq. (\ref{H-eq}) the only free parameter is the mass of black hole $M$
which is fixed from the start. Therefore, as long as classical physics are concerned,  the value of $H$ is fixed and inflation is eternal as seen from the conformal diagram in Fig. \ref{setup-fig}. 
However, it is known that a black hole is not an idle classical object. Indeed, it is  the source of black body radiation via Hawking radiation \cite{Hawking:1974sw, Hartle:1976tp} with the characteristic temperature 
\ba
T_H = \frac{1}{8 \pi G M} = \frac{1}{4 \pi  r_S} \, .
\ea
While it radiates with temperature $T_H$, it loses mass by the amount $\Delta M $ and correspondingly the radius of its event horizon  reduces by $\Delta  r_S / r_S = -\Delta M/M$. As it radiates further in subsequent steps, the radius of its event horizon becomes smaller and the black hole gets hotter. This pictures continues till black hole radiates its entire mass at a time scale of order $M^{3/}M_P^4$. Of course, towards its final stage of evaporation in which the radius of event horizon approaches $\ell_P$ the above picture may not hold and the  quantum gravity effects  are expected to play important roles. However, for our purpose the important point is that the black hole spends most of its semi-classical life time during the stage when $M$ is near its original value so the ratio  $\Delta  r_S/  r_S$ is small. 

The rate of change in black hole mass via Hawking radiation is given 
by \cite{Page:1976df}
\ba
\label{m-rate}
\frac{d M}{d \bar t} =  -\frac{\alpha M_P^4}{M^2} \, ,
\ea
in which $\alpha$ is a constant  determined by the number of all  massless degrees of freedom that the
black hole decays to. For the SM degrees of freedom at the temperature around the Higgs 
mass 125 $GeV$, we have $\alpha \sim 2$. However, the value of $\alpha$ can be significantly larger than unity 
for models beyond SM and it may be reasonable to 
take  $\alpha \lesssim 10^{2}$ if black hole decays at and above the GUT scale.   The above equation can be integrated to obtain the lifetime of black hole as measured from an observer far away from black hole 
\ba
\label{t-BH}
\bar t_{BH} \simeq \frac{M^3}{3 \alpha M_P^4} \, .
\ea  
As mentioned before, this means that the life time of black hole before complete evaporation scales like $M^3/M_P^4$. 

Now back to our first question as how inflation ends in this setup. Based on above discussions we argue that  the mechanism of end of inflation is provided by the Hawking radiation. Upon emitting thermally with the black body temperature $T_{H}$, the mass of black hole changes by $\Delta M $ and   the radius of its event horizon shrinks by $\Delta r_S$, causing a change in  Hubble expansion rate by $\Delta H$. The question is how slow this process of black hole evaporation is so inflation lasts for long enough period, say at least for 60 e-folds, to solve the flatness and the horizon problem. Intuitively speaking the mechanism of black hole evaporation is a very slow process so inflation can continue for a long period. The reason is that the mechanism for black hole evaporation is a quantum-thermal effect so it should be inefficient  for our black hole which is more or less a classical object. Note that we have assumed $M\gg M_P$ to trust classical general relativity analysis. For black hole which are significantly more massive than $M_P$ it will take a long time before 
the black hole evaporate from the quantum Hawking radiation.  Based on this argument, the number of e-folds of inflation is expected to be a positive power of $M_P/M$ because $M$ is the only free parameter of the theory. Below we argue that indeed the number of e-folds scale like $(M_{P}/M)^{2}$.

Let us define the parameter $\epsilon$ in our setup, similar to slow roll parameter in conventional models of inflation,  as the fraction change in $H$ in one Hubble expansion time $\Delta t =H^{-1} $
\ba
\label{epsilon}
\epsilon = -\frac{\Delta H}{H} \, .
\ea
As in  slow-roll models, inflation in our setup last long enough if $\epsilon $ is much less than unity while  inflation ends when $\epsilon \sim 1$. To calculate $\epsilon$ in the black hole side, we have to relate the cosmic time $\Delta t$ to the time measured on the black hole side which is $\Delta \bar t$ as defined in
Eq. (\ref{BH-metric1}).  For this purpose, note that from the relation $H= 1/r_S$ we obtain $\Delta H/H = - \Delta r_S/r_S$. So if $\Delta H/H$ is interpreted as the fractional change in $H$ in one Hubble expansion time, then $ \Delta r_S/r_S$ is interpreted as the fractional change in $r_S$ in the time scale $\Delta \bar t= r_S$. This is  consistent with the intuition that a time scale associated with the horizon of dS space is mapped to the 
time scale on the black hole side of order $r_S$, as one expects from the relation $r_S =H$. 

Now having obtained $\Delta H/H = - \Delta r_S/r_S$ with $\Delta \bar t= r_S$ as the corresponding time scale in the black hole side, we can use the differential relation of Hawking radiation Eq. (\ref{m-rate})
to calculate $\epsilon$. Specifically, using $r_S= 2 G M$,  so $\Delta r_S= 2 G \Delta M$,  and taking $\Delta \bar t= r_S$,  Eq. (\ref{m-rate}) yields 
\ba
\label{epsilon-result}
\epsilon = -\frac{\alpha}{4 \pi} \left( \frac{M_P}{M} \right)^2 = - 4 \pi \alpha \left( \frac{H}{M_P} \right)^2 \, .
\ea

Note the important effects that the change in $H$ is positive, i.e.  $\epsilon$ is negative. This is opposite to conventional models of inflation  in which 
the weak energy condition $\rho+ P \ge 0$ requires that $\dot H <0$. However, in our case the change 
in $H$ is induced via Hawking radiation which is a quantum effect. It seems that quantum effects such as
Hawking radiation can naturally provide a mechanism to violate null energy condition in cosmology. 
This can play crucial roles in other models of alternative to inflation such as bouncing cosmology. 
Indeed, the violation of the null energy condition was a challenging issue for models of bouncing cosmology  which require $\dot H <0$ during the period of bounce. 

Having obtained $\epsilon$ we can obtain an estimate of the total number of e-foldings $N$. Roughly speaking one expects that $N \sim 1/|\epsilon|$. Using Eq. (\ref{epsilon-result}) we obtain
\ba
N \sim  \frac{1}{4 \pi \alpha} \left( \frac{M_P}{H} \right)^2 \, .
\ea
This shows that there is no shortage in getting enough number of e-foldings. For example, taking $H/M_P \sim 10^{-5}$ and $\alpha \sim 10$, the above estimation  yields $N \sim 10^{8}$. Of course, observationally we only require the last 60 e-folds of inflation to solve the flatness and the horizon problem. We note that the larger is the value of $\alpha$, the smaller is $N$. This is because a large value of $\alpha$ corresponds to more channels that the black hole can decay to. If we allow for significantly many new degrees of freedom  in some beyond SM theories, such as grand unification or supersymmetric theories, then the black hole can decay to these new degrees of freedom so its lifetime becomes shorter and hence we obtain a lower value for $N$. 

As mentioned before, in our picture inflation ends via Hawking radiation when $\Delta H/H \sim 1$, i.e. when $|\epsilon| \sim 1$. This corresponds to scale $H \sim M_P$. Of course in this limit, we are far from
semi-classical description of the black hole physics and a full quantum gravity study is required. At this stage, the interior of black hole is very hot with all relativistic particles being excited. This sets  the reheating and  the origin of hot big bang cosmology inside the quantum  primordial black hole. 

We comment that in the above discussions we have neglected the deflation phase, i.e. the period 
$\eta < -\pi/2 $ in Fig. \ref{setup-fig}.  The main focus of our discussions here is the final stage of inflation for the period $-\pi/2<  \eta < \eta_f$. Of course, the period of deflation and bouncing cosmology may be interesting by its own right. It would be interesting if a period of bounce in this setup without reaching the final stage of inflation can solve the flatness and the horizon problem as an alternative to inflation \cite{Khoury:2001wf, Steinhardt:2001st, Ijjas:2013vea}. 

Before ending these discussions we note a curious effect in this setup. In our setup where $H= r_{S}^{-1}$ the black hole temperature in terms of $H$ is given by  $T_{H}= H/4 \pi$.  Curiously, this value of the black hole temperature  is by a factor $1/2$ smaller than the  temperature associated with the cosmological horizon of a dS background, $T_{dS}$,  calculated by Gibbons and Hawking \cite{Gibbons:1977mu}:  $T_{dS} = H/2 \pi$. This means that in our setup  the dS background is always hotter than the black hole $T_{dS}= 2 T_{H}$. We imagine that the black hole and the dS backgrounds are two separate thermal  systems which are in contact at the surface of black hole. Since $T_{dS} > T_{H}$, one may expect that  the black hole should get hotter for the two systems to  reach into thermal equilibrium. This is not in contradiction with Hawking radiation stating that as the black hole evaporates it gets hotter. Perhaps the system never reach a true thermal equilibrium since at each instant $T_{dS}= 2 T_{H}$. The distribution of energy and pressure on the surface of matching $\Sigma$ may prevent the two thermal systems to reach into a thermal equilibrium.

\subsection{Perturbations }
\label{perturbations}

After presenting our background inflation, now comes the important question how perturbations are generated in this setup to seed the perturbations in CMB and large scale structure.  As usual we look for scalar and tensor perturbations. The nature of tensor perturbations is easy to understand.  In conventional models of inflation, the tensor perturbations directly probe
the energy scale of inflation, i.e. the value of $H$ during inflation. The tensor perturbations are the perturbations in the
fabric of space-time without relying on the details of mechanism of inflation such as the slow-roll parameters. Since the tensor perturbations are easy to follow, here we start with the brief analysis of tensor perturbations.

\subsection{Tensor Perturbations}
\label{tensor}

The tensor perturbations of the metric are characterized  by \cite{Weinberg:2008zzc,Baumann:2009ds}
\ba
\label{tensor-metric}
ds^2 = - dt^2 + a(t)^2 (\delta_{ij} + h_{ij} ) dx^i dx^j \, .
\ea
The perturbations $h_{ij}$ are transverse and  traceless, $h_{ii}= \partial_i h_{ij}=0$,  in which it is understood that the indices are raised and lowered with the  flat metric $\delta_{ij}$. After imposing these conditions, only two degrees of freedom remain which correspond to two polarization modes of gravitational waves. 

Plugging the tensor perturbations into the gravitational action, for each polarization,  we obtain 
\ba
\label{S-h}
S_{h}= \frac{M_P^2}{8} \int d \eta d^3 x a^2 \left[ (h'_{ij})^2 - (\nabla h_{ij})^2  
\right]  \, ,
\ea
in which $\eta$ is the conformal time related to cosmic time via $d \eta = dt/a(t)$.  The above action has the simple form of a massless scalar field in the dS background. It is well-known that the amplitude of massless fields in dS background is $H/2\pi$.  Correspondingly, the power spectrum  of tensor perturbation has the universal form
\ba
\label{h-power}
{\cal P} _{h} = \frac{2}{\pi^2} \frac{H^2}{M_P^2} \Big |_{k =a H} \, ,
\ea
in which the value of $H$ is  calculated at the time of horizon crossing for the mode $k$ when $k =a H$.

Having presented the power spectrum of tensor perturbations, it is important to understand the quantum origin of these perturbations in both dS and and black hole sides. Let us start from the dS side.  We can obtain a good order of magnitude estimate for the amplitude of these perturbations as follows.  Let us look at the gravitational action
 \ba
 \label{S-gr}
S_{gr} = \frac{M_P^2}{2} \int d^4 x \, \sqrt {-g}  R  \, ,
\ea
in which $R$ is the Ricci scalar. The uncertainty principle implies that the space is full of quantum fluctuations. As an estimation of the amplitude of quantum fluctuations of the spacetime, suppose the relativistic length scale of these perturbations are at the order  $L$. Then the quantum effects are important if the action becomes at the order $\hbar =1$. Plugging these order
of magnitude estimations for the tensor perturbation in our gravitational action  Eq. (\ref{S-gr})  yield
$\Delta S_{gr} \sim L^4 M_P^2 L^{-2} = M_P^2 L^2$. Requiring that $\Delta S_{gr} = \hbar =1$, we obtain  $L \sim 1/M_P$.  Now let us see what this implies for the amplitude of quantum tensor perturbations $h_{ij}$. Using the metric Eq. (\ref{tensor-metric}), the  tensor perturbations cause the ripple in space with the physical length  $L \sim a  | {\bf x } | h_{phy}$ in which $h_{phy}$ means the amplitude of physical tensor perturbations (without caring about the polarization and indices). 
For the modes at the moment of horizon crossing $ a  | {\bf x } | \sim 1/H$ which is the physical length of dS geometry. Now, requiring that $L \sim 1/M_P$
as just obtained above, we obtain 
\ba
\label{h-amplitude}
h_{phy} \sim \frac{H}{M_P} \, .
\ea
This sets the amplitude of tensor modes.  Indeed, this is well consistent with the exact result for the power spectrum Eq. (\ref{h-power})
in which ${\cal P}_h \sim h_{phy}^2$ yielding $h_{phy} \sim H/M_P$ as given in Eq. (\ref{h-amplitude}). 

If we translate this back into our black hole picture, the tensor
perturbations are associated with the quantum fluctuations of the event horizon. In the black hole side there is the vast machinery of the black hole quasi normal modes in which the classical 
scalar, vector and tensor perturbations of black hole spacetime are  studied systematically \cite{Kokkotas:1999bd, Nollert:1999ji}. In principle, one can use these technology to make the link between our tensor perturbations in dS background and the corresponding tensor perturbations in black hole side precise. This is an interesting question, but it is beyond the scope of the present work. Instead, we follow our heuristic reasoning as used above to demonstrate that the above picture in dS side is quite consistent with its dual interpretation  in black hole side.  

At the classical level, the space on the black hole side is empty so $S_{gr}=R=0$. 
However, employing the uncertainty principle, we expect the quantum fluctuations of  spacetime to
cause ripples on the surface of event horizon. Following the same logic as in the dS case,  the length scale of quantum fluctuations which changes the gravitational action by $\hbar$ is obtained to be 
$L \sim 1/M_P$. After turning the tensor perturbations on the black hole background, the physical length is
given by $L \sim  r_{S} h_{ij}$ so setting $L \sim 1/M_{P}$ implies that $h_{{phy}} \sim 1/ r_{S} M_{P} \sim H/M_{P}$ which is in agreement with the result obtained in the dS side. 

The morale of the above heuristic presentation is that the quantum fluctuations of the tensor perturbations are sourced at the horizon of dS background or on the surface of event horizon of black hole 
with the amplitude $h_{phy}\sim H/M_P$ and on the  length scale $ \ell_P$. The conclusion that the quantum gravitational fluctuations of the black hole horizon  is the source of
the cosmological tensor perturbations is extended to  scalar perturbations which will be studied next. 

\subsection{Scalar Perturbations}
\label{scalar}

Here we study scalar perturbation in this setup. As just mentioned above, the cosmological scalar perturbations are associated with the quantum ripples at the surface of black hole event horizon. As such, from the above arguments 
for the tensor perturbations,  we expect that the length scale of quantum scalar perturbations on the surface of black 
horizon to be $\ell_{P}$ with the amplitude $H/M_{P}$. Consequently, the amplitude of curvature perturbations 
${\cal R}$ is expected to be ${\cal P}_{\cal R} \sim H^{2 }/M_{P}^{2}$. This also sets the amplitude of the observed COBE normalization for the temperature perturbations on CMB map ${\cal P}_{\cal R}^{1/2} \simeq \left( \delta T/T\right)_{{CMB}} \simeq 10^{-5}$,  providing  the rough estimate  $ H/M_{P} \sim 10^{-5}$.  

After this rough estimate for the amplitude of scalar perturbations, we parametrize  the scalar perturbation as
\ba
\label{R-beta}
{\cal R} \equiv \frac{\beta H}{M_{P}} \, .
\ea
Our goal here is to find the order of magnitude of  $\beta$.

Happily the question of vacuum fluctuation of black hole's event horizon has been studied by York in \cite{York:1983zb}. In an attempt to provide a dynamical explanation for the origin of Hawking radiation, he studied the black hole metric perturbation undergoing quantum zero point (vacuum) fluctuations. The underlying roles are the uncertainty principle  and the equivalence principle. It is argued that the tides associated with the zero point fluctuations distort the timelike limit surface, the apparent horizon and the event horizon. It is argued that this distortion creates the quantum ergosphere and  the desired quantum radiance is driven by the ``curvature fluctuations'',  as governed by the equivalence principle. It is pointed out that the zero point fluctuations can not sharply localize on the horizon because this would force the amplitude of metric fluctuations at the horizon to become very large. Instead, the uncertainty associated with a metric fluctuation with frequency $\omega$ is spread over  a region with the wavelength $2 \pi/\omega$.  Below we briefly summarize the results in \cite{York:1983zb} relevant for our analysis. 

The metric perturbations in the advanced Eddington-Finkelstein coordinate is given by 
\ba
\label{metric-York}
ds^{2 } = - (1- 2 G m/r) d v^{2 } + 2 dv dr + r^{2 } d \theta^{2} + r^{2} \sin^{2} \theta d \phi^{2} \, ,
\ea
in which 
\ba
\label{m-York}
m= m(v, \theta) = M + \sum_{\ell} ( 2 \ell + 1) \epsilon_{\ell } M \sin (\omega_{\ell} v)  q_{\ell} (\theta) \, .
 \ea
Here $M$ is the mass of the black hole in the absence of perturbations, $\ell$ is the usual angular momentum index labeling each mode and $q_{\ell}(\theta) = (2 \ell + 1)^{1/2} P_{\ell} (\cos \theta) $,  in which $P_{\ell}$ is the Legendre polynomial  determining the shape of black hole distortion. Finally, $\omega_{\ell}$ represents the resonant or ringing mode frequencies for each $\ell$.

After imposing the minimum uncertainty relation between the physical metric perturbations $\Delta g$ and its conjugate momentum, $\epsilon_{\ell}$ are obtained to be
\ba
\epsilon_{\ell}^{2 } = \frac{\hbar}{M^{2} } \Bigg[ \frac{3}{2} \left( \frac{ \sigma_{\ell} }{\pi} \right)^{3} 
\frac{1+ 16 \sigma_{\ell}^{2}}{\ell^{2} (\ell +1)^{2} + 16 \sigma_{\ell}^{2}}
\Bigg] \, ,
\ea
in which $\sigma_{\ell} \equiv G M \omega_{\ell}$ is the dimensionless frequency. For the first few $\ell$, the values of
$\sigma_{\ell}$ are $\sigma_{2 } \simeq 0.37, \sigma_{3} \simeq 0.60, \sigma_{4}\simeq 0.8, \sigma_{5} \sim 1.01$
and $\sigma_{6} \sim 1.21$. Of all $\ell$, only the quadrupole mode $\ell=2$ satisfies $\omega_{2}<  1/(2 G M)$, i.e. of all $\sigma_{\ell}$ only $\sigma_{2} < 0.5$. It is argued that the short wavelength modes act incoherently inside a quantum-mechanically defined volume of the system, in our case the radius of the event horizon. Therefor, one can neglect the contributions of modes beyond $\ell=2$ for which $\sigma_{\ell } >0.5$.  With this reasoning and other physical arguments, it is concluded that only the quadrupole mode is relevant in the metric perturbations. In other words, most of the gravitational radiation occurs at 
 $\ell=2$ in which the modes of $\ell=2$ propagate in an essentially free radial manner towards the distant observer. 

With $\ell=2$ as the dominant mode of perturbations, the change in the mass from Eq. (\ref{m-York}) is obtained to be $\Delta M \sim \epsilon_{2 } M \simeq 1.8 \sqrt{\hbar/G}$. More specifically, the characteristic rms fluctuations 
of the `` irreducible'' mass of the physical event horizon is obtained to be
\ba
\label{delta-M}
\Delta M_{EH} &\equiv& \sqrt{5} \Big[  \langle M_{EH}^{2 } \rangle - \langle M_{EH} \rangle^{2}  \Big]^{1/2} 
\nonumber\\
&\simeq& 1.18\times 10^{-2} \sqrt{\hbar/G} \simeq  9 \times 10^{-2}  M_{P} \, .
\ea
Note that the factor $\sqrt 5$ comes from the contributions of $2 \ell+1 =5$ independent modes associated with $\ell=2$. Also note that the averaging above is over one period of time $\tau= 2 \pi/\omega_{\ell}$ and also over the surface of the two-dimensional sphere. 

The above value of $\Delta M_{EH}$ sets the time scale of vacuum fluctuations on the surface of black hole horizon. Taking the corresponding uncertainty in time scale as $\Delta t \sim 1/ \Delta E \sim 1/\Delta M_{EH}$, we obtain 
\ba
\label{delta-t}
\Delta t \simeq 17 M_{P}^{-1} \, .
\ea
This is consistent with our previous rough estimate that the length (time) scale associated with the vacuum fluctuations  of metric is at the order $L \sim  M_{P}^{-1} $. However, the results in 
\cite{York:1983zb} enable us  to find the numerical pre-factor relating $\Delta t$ to $M_{P}^{-1}$ which would be necessary to determine the parameter $\beta$ defined in Eq. (\ref{R-beta}). 

 Having obtained the time scale of curvature fluctuations on the horizon given in Eq. (\ref{delta-t}) we are able to estimate curvature perturbations ${\calR}$. Going to flat gauge, ${\calR}$ is given by ${\calR}= - H \Delta t$.
 With the  estimation for $\Delta t$ given in Eq. (\ref{delta-t}) we obtain
 \ba
 \label{R-eq}
 {\cal R} \simeq \frac{17 H}{M_{P}} \, ,
 \ea
which fixes our numerical parameter $\beta \simeq 17$. 

Using the COBE normalization ${\calR} \simeq 2\times 10^{-5} $, the scale of inflation in our model is fixed to be
$H \simeq 10^{-6} M_{P}$. It is interesting that the scale of inflation is uniquely determine by the scalar  perturbations. This should be compared with the situation in models of slow-roll inflation in which the scale of inflation can not be determined uniquely by  curvature perturbations. In these models ${\calP_{\calR}} = H^{2}/8 \pi^{2} \epsilon  M_{P}^{2}$ or ${\calR} \simeq H/\sqrt{8 \epsilon} \pi M_{P}$ so the scale of inflation can be fixed modulo the unknown slow-roll parameter $\epsilon$.  Now comparing our result for ${\calR}$ with the predictions of slow-roll models, we make the identification $\beta \leftrightarrow 1/\sqrt{8 \epsilon } \pi$.  

Having obtained both the scalar and the tensor power spectra, now we can look at their ratio. Defining their ratio as usual via $r \equiv 2 {\calP}_{h}/{\calP}_{\calR}$ we obtain
\ba
r = \frac{2}{\beta^{2} \pi^{2}} \simeq 7\times  10^{-4}  \, .
\ea
This value for the parameter $r$ is too small to be detected by the current CMB observations. It is interesting to see if the next generation of CMB observations can detect the above value of  $r$. We emphasis that the analysis presented above provide the order of magnitude of the parameters $\beta$ and $r$. A careful analysis of scalar curvature perturbations are necessary to fix the parameter $\beta$. Even a factor $\pi$ uncertainty  in our heuristic estimate of $\beta$ based on the results of \cite{York:1983zb} can enhance the value of $r$ by $\pi^{2 } \sim 10$. Therefore, a more precise analysis of curvature perturbations beyond what presented here is necessary.

Finally, we look at the spectral index $n_{s}$ defined via $n_{s}-1 = d \ln {\calP}_{\calR}/d \ln k$. Following the usual 
method and noting that all quantities are calculated at the time of horizon crossing when $k =a H$ we obtain
\ba
\label{ns}
n_{s} -1 \simeq -2 \epsilon + \frac{2 \dot \beta}{H \beta} \, ,
\ea
in which $\epsilon =-\Delta H/H \simeq -\dot H/H^{2}$. Note that in obtaining the above result, we have allowed for the possibility that the parameter $\beta$ may have a mild
 scale-dependence. In other words, the parameter $\beta$ for the vacuum fluctuations of the black hole horizon may have a mild dependence on the mass of black hole, $\beta = \beta(M)$. 
 
 In our model, $\epsilon \sim - H^{2}/M_{P}^{2 } \sim - 10^{-12}$ so 
 if the parameter $\beta$ has no running then $n_{s}-1$ is practically scale invariant. However, the current observations by the Planck team has rule out a scale invariant power spectrum beyond 5$\sigma$ confidence level
 \cite{Ade:2015lrj}. Therefore, it is desirable to obtain a red-tilted power spectrum with $n_{s} <1$. Of course, a mild scale-dependence in the parameter $\beta$ may come to rescue. This is another reason why the scalar perturbations in this setup has to be performed more carefully to see if the parameter $\beta$ can have any scale dependence to yield a red-tilted power spectrum. 
 
 We comment that so far we have concentrated on the simple setup of FRW Universe inside the Schwarzschild black hole. Alternatively, one can consider the case of a charged black hole with the Reissner-Nordstrom metric as mentioned at the end of Section \ref{setup}. In this case, the energy density and the Hubble expansion rate during inflation have extra dependence on the electric charge of black hole, $Q$,  as given in Eq. (\ref{rho-Q}). If the discharge of black hole from its electric charge via Hawking radiation  is faster than losing its mass, then 
 $\epsilon $ can be much larger than its value in the Schwarzschild black hole case. Also note that since $Q$ contributes negatively in $H$,  its discharge yields a positive contribution to $\epsilon$ so a red-tilted power spectrum may be achieved from Eq. (\ref{ns}).  Finally, one can consider the more complicated case of the charged Kerr metric with the three parameters, mass, charge and the angular momentum $J$. The possibility of angular momentum is somewhat intriguing. Intuitively speaking, the orientation  of angular momentum defines a preferred direction so inflation inside the Kerr black hole may be in the form of anisotropic inflation \cite{Watanabe:2009ct, Emami:2010rm}. This may result in quadrupole statistical anisotropy in curvature power spectrum.

 
 \section{Summary and Discussions}
 \label{summary}
 
 In this work we have entertained the possibility that our primordial Universe was inside a primordial black hole. This is motivated from the fact that there are curious similarities between the physics of black hole and the dS spacetimes. The fact that both spacetimes have horizon is the key for our setup. The interior of the event horizon of the black hole is mapped to a dS background while the two spacetimes join smoothly at their boundaries. This allows us
 to provide a one to one map with the known physics of black hole to the corresponding dynamics in the dS side.

 The above picture  provides an intriguing mechanism of inflation without invoking  the inflaton field. 
 It is argued that the built-in Hawking radiation for the evaporation of black hole provides the natural mechanism to terminate inflation. Due to the slow quantum process of Hawking radiation, inflation proceeds for a long period. The model typically predicts  $N \sim (M_{P}/H)^{2}$ e-folds of inflation.  The reheating and the onset of hot big bang cosmology in this setup happens when black hole radiates all its energy to relativistic particles inside the (remnant) quantum black hole. 
 The final stage of inflation in which the black hole mass and temperature approach the Planck energy scale is not well-understood. This is the limit that the quantum gravity effects can play important roles and the physics beyond SM are quite important. The decay rate of the black hole, determined by the parameter $\alpha$ in Eq. (\ref{m-rate}), depends strongly on the number of all massless particles that the black hole can decay to. At the final stage of its evaporation, the numerous degrees of freedom of 
 the physics beyond SM are expected  to be massless so the parameter $\alpha$ can be many orders of magnitude larger than unity. 
 
In the simplest case that the black hole decays only to SM particles, the faction of decay channels are determined by the spin of particles, $s=0, 1/2, 1$ and $s=2$ for two modes of gravitons. The fraction of energy deposited to gravitons is about few percents \cite{Page:1976df}. Therefore, our model predicts the existence of background gravitations with a fraction of the energy of background photons. Note that this is different than the gravitational wave perturbations on CMB which was calculated in Section  \ref{tensor}. 
It would be interesting to see if a few percent of background radiation energy density  in the form of isotropic and homogeneous  gravitons particles can be detected observationally.  

The universality of the decay of black hole via Hawking radiation implies that the perturbations to be adiabatic. Since the decay rate of the black hole to each species (baryons, leptons, photons etc) are uniquely determined via their spins, it is expected that the perturbations in the energy density of each species to follow its background energy density, i.e. $\delta \rho_{b}/\rho_{b} = \delta \rho_{\gamma}/\rho_{\gamma}$ etc. As a result,   the adiabaticity of the CMB perturbations is a natural outcome of this model. In addition, since the background is nearly dS with no significant 
interactions, it is expected the perturbations to be nearly Gaussian. Putting it another way, the interactions involved in the system are sourced from gravitational back-reactions, so motivated by the analysis of \cite{Maldacena:2002vr}, we expect the system to generate no large amount of non-Gaussianity. However, towards the end of black hole's life, the quantum gravitational effects become significant. Therefore, the simple picture of small gravitational back reactions is modified towards the final stage of inflation and small-scales non-Gaussianities may be generated towards the final stage of inflation. 

The model can naturally incorporate the existence of some massive species beyond the SM sector as the source of dark matter and mechanism of baryon asymmetry. For example, as a possible mechanism of baryon asymmetry, at the time of black hole decay it can channel a fraction of its energy to a particle with the mass of GUT scale which goes out of thermal equilibrium afterwards. If this particle decays to baryons and anti baryons asymmetrically with different decay rates, it can generate the seeds of baryon asymmetry  as in mechanism of delayed decay of heavy  particles \cite{Weinberg:2008zzc}.

In this work we did not speculate on the origin of the primordial black hole itself. It may have been created spontaneously from vacuum. Alternatively, it might have existed eternally in the past. In addition, we have considered the ideal situation that the black hole has reached its final mass $M$, i.e. it is not accreting matter from outside. We have assumed that the interior allows an isotropic and homogenous solution. These assumptions are far from obvious if the black hole is not in its final configuration and matter are absorbed into it. 

One drawback of our setup is that the matching conditions to glue a dS space to the interior of a black hole on the surface of event horizon is non-trivial. The actual matching requires a divergence in surface pressure \cite{Barrabes:1991ng, Poisson:1988wc}.  We have no mechanism at hand to resolve this issue. The solutions to this shortcoming may involve non-trivial quantum physics around the horizon of black hole. This is an open question which should be addressed for our setup of primordial inflationary Universe inside the black hole to be treated as a consistent scenario.  

Finally, the speculation that the primordial Universe was inside a single primordial black hole opens up the way for many more speculations! For example, if there are more than one primordial black hole, then inside each black hole there is an FRW Universe. This naturally suggest the emergence of ``multiverse'' in this picture. Another possible speculation is that our present Universe itself  is inside a huge black hole. This may provide some clues  for the  cosmological constant problem. Indeed the  idea that our current Universe is inside a black hole 
was put forward in \cite{Easson:2001qf}  as a solution to the flatness and the horizon problem without relying on inflation.  Finally, the question of big bang singularity in our setup has been traded by the question of black hole singularity. But the possibility that the interior of black hole allows for a cosmological solution may suggest that 
there was no singularity of any type after all. This is in line with the  maximum curvature proposal suggested 
in \cite{Poisson:1988wc, Frolov:1989pf, Frolov:1988vj}.


\vspace{0.7cm}

{\bf Acknowledgments:} I would like to thank Robert Brandenberger and Misao Sasaki for useful discussions and correspondences. I am grateful to  M. A. Gorji, Sadra Jazayeri and  Javad Taghizadeh Firouzjaee for numerous discussions and clarifications. 

\appendix


\section{Matching surface in static coordinate}
\label{static}

Here we provide some sketches on the difficulty of matching a dS solution to the event horizon of the black hole. For this purpose we use the static coordinate
\ba
ds^{2} = -f(r) d\, t^{2 } + \frac{dr^{2}}{f(r)} + r^{2} d \Omega^{2}
\ea
in which $f(r)$ for two sides of spacetime is defined via
\ba
\label{f-form}
f(r) =   \Bigg{\{} 
\begin{array}{c}
1- H^{2} r^{2} \quad  0\leq r \leq r_{0}\\
1- \frac{2 G M}{r} \quad  r \ge r_{0} 
\end{array} 
\ea
The surface of matching $\Sigma$ is located at  $r=r_{0}$. For the moment we have not specified the position 
of $r_{0}$ but our goal is to consider the case where $r_{0} = 2 G M = H^{-1}$. 

The continuity of the metric requires that $H^{2 } r_{0}^{2} = 2 G M/r_{0}$ so $r_{0} = 2 G M/H^{2} = r_{S}/H^{2}$ in which $r_{S} = 2 GM$ is the black hole's event horizon.

Now we investigate the jumps in extrinsic curvature. The unit normal vector to the surface of matching is
\ba
n_{\mu} = (0, \frac{1}{\sqrt{f(r_{0})}}, 0, 0 )  \, .
\ea
Denoting the components of $K_{\alpha \beta}$ on the black hole's side and on the dS side respectively by 
$K^{>}_{\alpha \beta}$ and $K^{<}_{\alpha \beta}$ we have 
\ba
{ K^{>} }^{t}_{ t}  &=& -\frac{f'}{2 f}  = \frac{G M}{r_{0}} \left( 1- \frac{2 G M}{r_{0}}\right)^{-1/2} \\
{K^{>} }^{ \theta}_{ \theta}  &=& - \frac{1}{r_{0}}  \sqrt{1- \frac{2 G M}{r_{0}}} \\
{K^{>} }^{ \phi}_{ \phi} &=&  {K^{>} }^{ \theta}_{ \theta}\, ,
\ea
and
\ba
{ K^{<} }^{t}_{ t}  &=& -\frac{f'}{2 f}= - H^{2 } r_{0}   \left( 1- H^{2} r_{0}^{2} \right)^{-1/2}\\
{K^{<} }^{ \theta}_{ \theta}  &=& - \frac{1}{r_{0}}    \sqrt{ 1- H^{2} r_{0}^{2}} \\
{K^{<} }^{ \phi}_{ \phi} &=&  {K^{<} }^{ \theta}_{ \theta} \, .
\ea

The discontinuities in extrinsic curvature should be balanced by  characteristic integrals of  energy density over the surface of matching $\Sigma$ at the position $r_{0}$. If we take $H^{-1} < r < r_{0}$ then imposing 
the matching condition one arrives at the solution obtained in \cite{Frolov:1989pf, Frolov:1988vj}. 
However, if we consider the extreme case $r_{0 } = 2 GM = H^{-1}$ then we see the appearance of pathologies. 
More specifically, ${K }^{ \phi}_{ \phi}$ and $  {K}^{ \theta}_{ \theta}  $ are well-defined on both sides and indeed are continues if $r_{0 } = 2 GM = H^{-1}$. However, ${ K }^{t}_{ t}$  diverges at this point. The divergence in 
${ K }^{t}_{ t}$ is easily traced to the divergence in unit normal $n^{\mu}$ in which at horizon $f(r_{S})=0$. Of course this can be an artifact of coordinate singularity associated with the Schwarzschild metric on the surface of event horizon so one has to use the non-singular extension of the $(r, t)$ coordinate such as the Kruskal-Szekeres coordinate. However, the difficulty in matching a dS solution to a black hole solution at the surface of event horizon 
in the form of Eq. (\ref{f-form}) is genuine and one may have  to consider an infinite  pressure on the surface of matching  \cite{Poisson:1988wc,  Barrabes:1991ng}.

{}


\begin{thebibliography}{}

\bibitem{Guth:1980zm}
  A.~H.~Guth,
  Phys.\ Rev.\ D {\bf 23}, 347 (1981).

\bibitem{Linde:1981mu}
  A.~D.~Linde,
  Phys.\ Lett.\ B {\bf 108}, 389 (1982).

\bibitem{Albrecht:1982wi}
  A.~Albrecht and P.~J.~Steinhardt,
  Phys.\ Rev.\ Lett.\  {\bf 48}, 1220 (1982).

\bibitem{Sato:1980yn}
  K.~Sato,
  Mon.\ Not.\ Roy.\ Astron.\ Soc.\  {\bf 195}, 467 (1981).

\bibitem{Starobinsky:1980te}
  A.~A.~Starobinsky,
  Phys.\ Lett.\ B {\bf 91}, 99 (1980).

\bibitem{Ade:2015lrj} 
  P.~A.~R.~Ade {\it et al.} [Planck Collaboration],
  arXiv:1502.02114 [astro-ph.CO].

\bibitem{Ade:2015xua} 
  P.~A.~R.~Ade {\it et al.} [Planck Collaboration],
  arXiv:1502.01589 [astro-ph.CO].


\bibitem{Planck:2013jfk} 
  P.~A.~R.~Ade {\it et al.} [Planck Collaboration],
  Astron.\ Astrophys.\  {\bf 571}, A22 (2014)
  [arXiv:1303.5082 [astro-ph.CO]].

\bibitem{Khoury:2001wf} 
  J.~Khoury, B.~A.~Ovrut, P.~J.~Steinhardt and N.~Turok,
  Phys.\ Rev.\ D {\bf 64}, 123522 (2001)
  [hep-th/0103239].

\bibitem{Steinhardt:2001st} 
  P.~J.~Steinhardt and N.~Turok,
  Phys.\ Rev.\ D {\bf 65}, 126003 (2002)
  [hep-th/0111098].

\bibitem{Ijjas:2013vea} 
  A.~Ijjas, P.~J.~Steinhardt and A.~Loeb,
  Phys.\ Lett.\ B {\bf 723}, 261 (2013)
  [arXiv:1304.2785 [astro-ph.CO]].

\bibitem{Brandenberger:1988aj} 
  R.~H.~Brandenberger and C.~Vafa,
  Nucl.\ Phys.\ B {\bf 316}, 391 (1989).

\bibitem{Nayeri:2005ck} 
  A.~Nayeri, R.~H.~Brandenberger and C.~Vafa,
  Phys.\ Rev.\ Lett.\  {\bf 97}, 021302 (2006)
  [hep-th/0511140].

\bibitem{Abbott:2016blz} 
  B.~P.~Abbott {\it et al.} [LIGO Scientific and Virgo Collaborations],
  Phys.\ Rev.\ Lett.\  {\bf 116}, no. 6, 061102 (2016)
  [arXiv:1602.03837 [gr-qc]].

\bibitem{Abbott:2016nmj} 
  B.~P.~Abbott {\it et al.} [LIGO Scientific and Virgo Collaborations],
  Phys.\ Rev.\ Lett.\  {\bf 116}, no. 24, 241103 (2016)
  [arXiv:1606.04855 [gr-qc]].

\bibitem{Hawking:1974sw} 
  S.~W.~Hawking,
  Commun.\ Math.\ Phys.\  {\bf 43}, 199 (1975)
  Erratum: [Commun.\ Math.\ Phys.\  {\bf 46}, 206 (1976)].

\bibitem{Hartle:1976tp} 
  J.~B.~Hartle and S.~W.~Hawking,
  Phys.\ Rev.\ D {\bf 13}, 2188 (1976).


\bibitem{Gibbons:1977mu} 
  G.~W.~Gibbons and S.~W.~Hawking,
  Phys.\ Rev.\ D {\bf 15}, 2738 (1977).

\bibitem{Padmanabhan}
T. Padmanabhan, `` Gravitation: Foundations and Frontiers,'' 2010. 


\bibitem{Sakharov:1966aja} 
  A.~D.~Sakharov,
  Zh.\ Eksp.\ Teor.\ Fiz.\  {\bf 49}, no. 1, 345
  [Sov.\ Phys.\ JETP {\bf 22}, 241 (1966)].

\bibitem{Poisson:1988wc} 
  E.~Poisson and W.~Israel,
  Class.\ Quant.\ Grav.\  {\bf 5}, L201 (1988).

\bibitem{Frolov:1989pf} 
  V.~P.~Frolov, M.~A.~Markov and V.~F.~Mukhanov,
  Phys.\ Lett.\ B {\bf 216}, 272 (1989).

\bibitem{Frolov:1988vj} 
  V.~P.~Frolov, M.~A.~Markov and V.~F.~Mukhanov,
  Phys.\ Rev.\ D {\bf 41}, 383 (1990).

\bibitem{Shen-Zhu}
W. Shen and S. Zhu, {\it Phys. Lett.} {\bf  126A 229} (1988). 

\bibitem{Sato:1981bf} 
  K.~Sato, M.~Sasaki, H.~Kodama and K.~i.~Maeda,
  Prog.\ Theor.\ Phys.\  {\bf 65}, 1443 (1981).

\bibitem{Maeda:1981gw} 
  K.~i.~Maeda, K.~Sato, M.~Sasaki and H.~Kodama,
  Phys.\ Lett.\ B {\bf 108}, 98 (1982).

\bibitem{Sato:1981gv} 
  K.~Sato, H.~Kodama, M.~Sasaki and K.~i.~Maeda,
  Phys.\ Lett.\ B {\bf 108}, 103 (1982).


\bibitem{Farhi:1986ty} 
  E.~Farhi and A.~H.~Guth,
  Phys.\ Lett.\ B {\bf 183}, 149 (1987).

\bibitem{Blau:1986cw} 
  S.~K.~Blau, E.~I.~Guendelman and A.~H.~Guth,
  Phys.\ Rev.\ D {\bf 35}, 1747 (1987).

\bibitem{Oshita:2016btk} 
  N.~Oshita and J.~Yokoyama,
  arXiv:1601.03929 [gr-qc].


\bibitem{Markov}
M.  A. Markov, Pis'ma Zh. Eksp. Teor. Fiz. {\bf 36, 214} (1982) [ Sov. Phys JETP Lett. 36, 265 (1982)]. 

\bibitem{Israel:1966rt} 
  W.~Israel,
  Nuovo Cim.\ B {\bf 44S10}, 1 (1966)
  [Nuovo Cim.\ B {\bf 44}, 1 (1966)]
  Erratum: [Nuovo Cim.\ B {\bf 48}, 463 (1967)].

\bibitem{Balbinot:1990zz} 
  R.~Balbinot and E.~Poisson,
  Phys.\ Rev.\ D {\bf 41}, 395 (1990).

\bibitem{Dymnikova:1992ux} 
  I.~Dymnikova,
  Gen.\ Rel.\ Grav.\  {\bf 24}, 235 (1992).

\bibitem{Dymnikova:1999cz} 
  I.~G.~Dymnikova,
  Phys.\ Lett.\ B {\bf 472}, 33 (2000)
  [gr-qc/9912116].

\bibitem{Barrabes:1991ng} 
  C.~Barrabes and W.~Israel,
  Phys.\ Rev.\ D {\bf 43}, 1129 (1991).

\bibitem{Gron}
O. Gron, {\it Lett. Nuovo Cimento } {\bf 44 177} (1985).  

\bibitem{Mukhanov:2005sc} 
  V.~Mukhanov,
  ``Physical Foundations of Cosmology''. 


 

\bibitem{Page:1976df} 
  D.~N.~Page,
  Phys.\ Rev.\ D {\bf 13}, 198 (1976).
  
\bibitem{Weinberg:2008zzc} 
  S.~Weinberg,
  ``Cosmology,''
  Oxford, UK: Oxford Univ. Pr. (2008) 593 p. 



\bibitem{Baumann:2009ds} 
  D.~Baumann, 
  ``Inflation,'' arXiv:0907.5424 [hep-th].

\bibitem{Kokkotas:1999bd} 
  K.~D.~Kokkotas and B.~G.~Schmidt,
  Living Rev.\ Rel.\  {\bf 2}, 2 (1999)
  [gr-qc/9909058].

\bibitem{Nollert:1999ji} 
  H.~P.~Nollert,
  Class.\ Quant.\ Grav.\  {\bf 16}, R159 (1999).

\bibitem{York:1983zb} 
  J.~W.~York, Jr.,
  Phys.\ Rev.\ D {\bf 28}, 2929 (1983).

\bibitem{Watanabe:2009ct}
  M.~a.~Watanabe, S.~Kanno and J.~Soda,
  Phys.\ Rev.\ Lett.\  {\bf 102}, 191302 (2009)
  [arXiv:0902.2833 [hep-th]].
  
\bibitem{Emami:2010rm}
R.~Emami, H.~Firouzjahi, S.~M.~Sadegh Movahed, M.~Zarei,
JCAP {\bf 1102 } (2011)  005.
[arXiv:1010.5495 [astro-ph.CO]].

\bibitem{Maldacena:2002vr} 
  J.~M.~Maldacena,
  JHEP {\bf 0305}, 013 (2003)
  [astro-ph/0210603].

\bibitem{Easson:2001qf} 
  D.~A.~Easson and R.~H.~Brandenberger,
  JHEP {\bf 0106}, 024 (2001)
  [hep-th/0103019].


\end{thebibliography}
\end{document}